\documentclass[journal,hideappendix]{vgtc}        
\usepackage{multicol}

\onlineid{9021}

\vgtccategory{Research}

\vgtcpapertype{application/design study}

\newcommand{\techName}[1]{\textit{PrettiSmart}}

\newcommand{\mO}[1]{Simulation Overview Module}
\newcommand{\mD}[1]{Simulation Detail Module}

\newcommand{\red}{\includegraphics[scale=0.15]{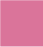}}%
\newcommand{\blue}{\includegraphics[scale=0.15]{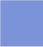}}%
\newcommand{\dark}{\includegraphics[scale=0.15]{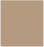}}%
\newcommand{\light}{\includegraphics[scale=0.15]{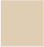}}%
\newcommand{\lred}{\includegraphics[scale=0.15]{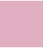}}%
\newcommand{\lblue}{\includegraphics[scale=0.15]{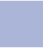}}%
\newcommand{\lgrey}{\includegraphics[scale=0.15]{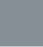}}%
\newcommand{\lgreen}{\includegraphics[scale=0.15]{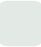}}%

\newcommand{\uptri}{\includegraphics[scale=0.2]{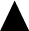}}%
\newcommand{\downtri}{\includegraphics[scale=0.2]{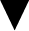}}%

\newcommand{\upin}{\includegraphics[scale=0.15]{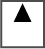}}%
\newcommand{\downin}{\includegraphics[scale=0.15]{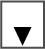}}%
\newcommand{\downout}{\includegraphics[scale=0.15]{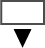}}%
\newcommand{\dline}{\includegraphics[scale=0.15]{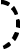}}%
\newcommand{\hdline}{\includegraphics[scale=0.15]{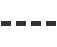}}%
\newcommand{\fout}{\includegraphics[scale=0.15]{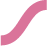}}%
\newcommand{\fin}{\includegraphics[scale=0.15]{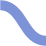}}%
\newcommand{\gcircle}{\includegraphics[scale=0.15]{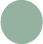}}%
\newcommand{\circleR}{\includegraphics[scale=0.15]{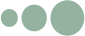}}%
\newcommand{\rborder}{\includegraphics[scale=0.15]{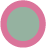}}%
\newcommand{\bborder}{\includegraphics[scale=0.15]{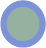}}%
\newcommand{\gtri}{\includegraphics[scale=0.15]{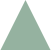}}%
\newcommand{\addr}{\includegraphics[scale=0.15]{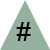}}%
\newcommand{\Oaddr}{\includegraphics[scale=0.15]{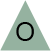}}%
\newcommand{\fcircle}{\includegraphics[scale=0.15]{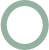}}%
\newcommand{\ftri}{\includegraphics[scale=0.15]{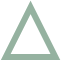}}%
\newcommand{\darkup}{\includegraphics[scale=0.15]{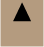}}%
\newcommand{\lightdowndown}{\includegraphics[scale=0.15]{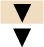}}%
\newcommand{\darkupdowndown}{\includegraphics[scale=0.15]{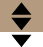}}%
\newcommand{\lightdown}{\includegraphics[scale=0.15]{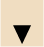}}%
\newcommand{\redcircle}{\includegraphics[scale=0.15]{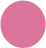}}%

\newcommand{\mod}[1]{\textcolor{black}{#1}}

\title{\techName{}: Visual Inter\underline{pret}a\underline{ti}on of \underline{Smart} Contracts via Simulation}
\author{%
  \authororcid{Xiaolin Wen}{0000-0002-8562-7640},
    \authororcid{Tai D. Nguyen}{0000-0003-2109-845X},
\authororcid{Lun Zhang}{0009-0002-9543-1964},
      \authororcid{Jun Sun}{0000-0002-3545-1392},
    and \authororcid{Yong Wang}{0000-0002-0092-0793}
}

\authorfooter{
  \item
  	Xiaolin Wen and Yong Wang are with Nanyang Technological University.
  	E-mail: xiaolin004@e.ntu.edu.sg, yong-wang@ntu.edu.sg. 
    Yong Wang is the corresponding author.
    
  \item Tai D. Nguyen and Jun Sun are with Singapore Management University.
  	E-mail: \{dtnguyen.2019, junsun\}@smu.edu.sg.
    
  \item
  	Lun Zhang is with GoPlus Security, China.
  	E-mail: allen@gopluslabs.io.

}

\abstract{%
Smart contracts are the fundamental components of blockchain technology. They are programs to determine cryptocurrency transactions, and are irreversible once deployed, making it crucial for cryptocurrency investors to understand the cryptocurrency transaction behaviors of smart contracts comprehensively. 
However, it is a challenging (if not impossible) task for investors, as they do not necessarily have a programming background to check the complex source code.
Even for investors with certain programming skills, inferring all the potential behaviors from the code alone is still difficult, since the actual behaviors can be different when different investors are involved.
To address this challenge, we propose \techName{}, a novel visualization approach via execution simulation to achieve intuitive and reliable visual interpretation of smart contracts.
Specifically, we develop a simulator to comprehensively capture most of the possible real-world smart contract behaviors, involving multiple investors and various smart contract functions. Then, we present \techName{} to intuitively visualize the simulation results of a smart contract, which consists of two modules:
The \mO{} is a barcode-based design, providing a visual summary for each simulation, and the \mD{} is an augmented sequential design to display the cryptocurrency transaction details in each simulation, such as function call sequences, cryptocurrency flows, and state variable changes. It can allow investors to intuitively inspect and understand how a smart contract will work.
We evaluate \techName{} through two case studies and in-depth user interviews with 12 investors.
The results demonstrate the effectiveness and usability of \techName{} in facilitating an easy interpretation of smart contracts.
}

\keywords{Smart Contract, Visualization, Simulation, Blockchain}

\teaser{
  \centering
    \setlength{\abovecaptionskip}{0.1cm}
  \setlength{\belowcaptionskip}{-0.2cm}
  \includegraphics[width=0.95\linewidth, alt={A view of a city with buildings peeking out of the clouds.}]{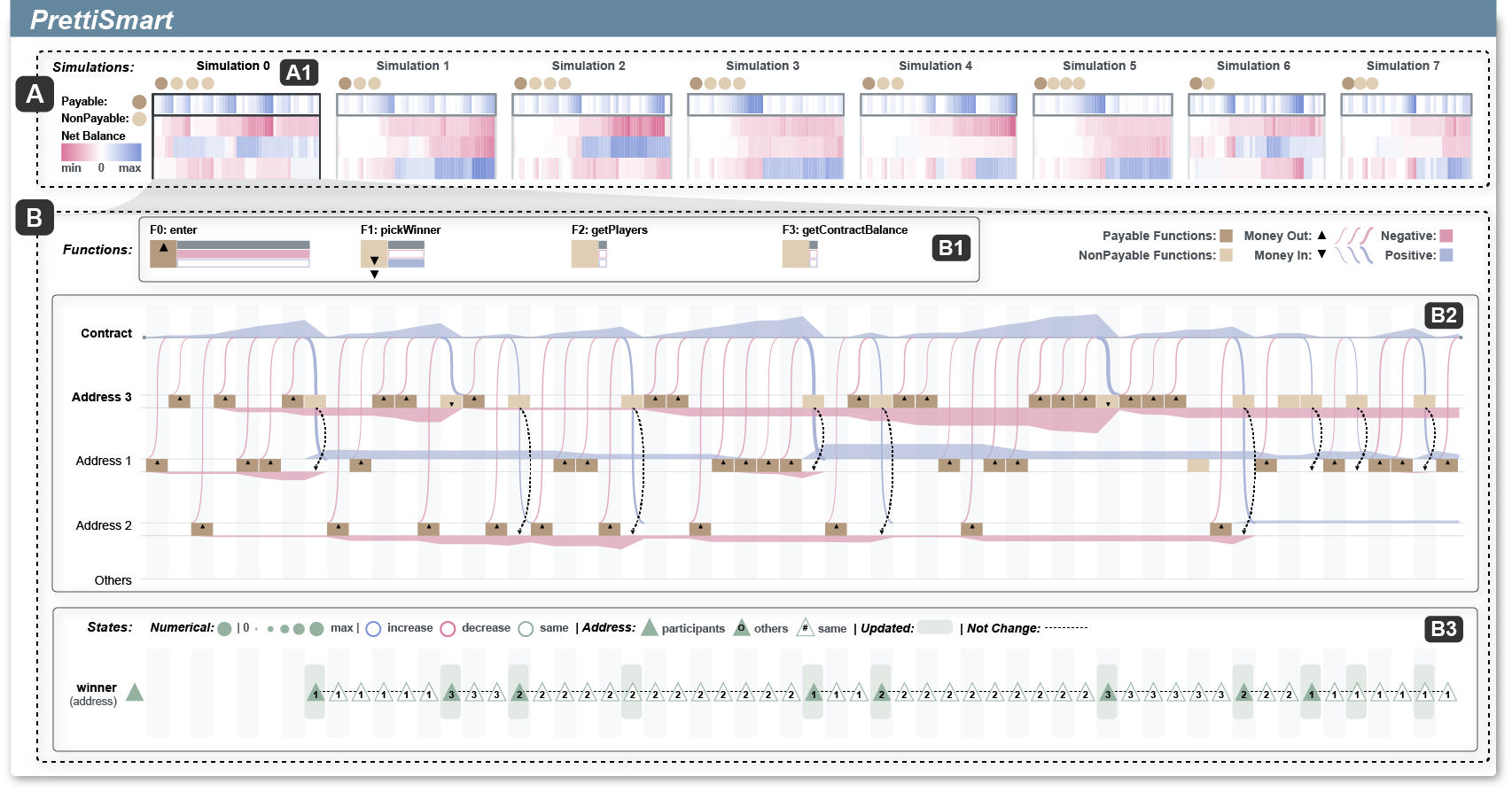}
  \caption{%
  The interface of \techName{} consists of a \textit{\mO{}} (A) to provide a visual summary for each simulation (A1) generated by our simulator and a \textit{\mD{}} (B) to show the details of each simulation, including the Function Summary (B1), the Function Call Details involving cryptocurrency flows and net balance changes (B2), and State Variable Changes (B3). In this figure, \techName{} provides a visual interpretation of a gambling smart contract, called ``Lottery", and helps investors intuitively understand the contract's behavior after each investment and how it selects and repays the winner.
  }
  \label{fig:teaser}
}

\graphicspath{{figs/}{figures/}{pictures/}{images/}{./}} 

\usepackage{tabu}                      
\usepackage{booktabs}                  
\usepackage{lipsum}                    
\usepackage{mwe}                       

\usepackage{mathptmx}                  

\begin{document}

\maketitle
\section{Introduction}
With the surging popularity of blockchain technology,
smart contracts, the essential component of blockchain, have recently gained increased attention~\cite{khan2021blockchain}.
\mod{Smart contracts are programs deployed on the blockchain, which will be executed when predefined conditions
are met or specific trigger events occur~\cite{zheng2020overview}.}
Due to their ability to provide high transparency without the need for trusted third parties, smart contracts are widely used in various blockchain-based applications, such as digital asset exchange, decentralized finance, health care, and supply chain management~\cite{huang2019smart,kushwaha2022ethereum}.
According to statistics from Etherscan~\cite{etherscan}, a popular blockchain explorer, the number of smart contracts deployed on the Ethereum blockchain has exceeded 65 million as of February 2024.
As reported by The Block~\cite{theblock}, a well-known blockchain information website, the total invested value in smart contract applications exceeds $\$$100 billion as of March 2024.

\mod{
Along with the wide usage of smart contracts in blockchain, there is an increasing necessity for investors to comprehend their functionality and assess possible risks before investing. Since the transactions recorded on the blockchain are irreversible, investors can not retrieve their money once a transaction is made~\cite{zheng2023securing}. 
In this paper, the term ``investors" refers to individuals planning to invest in projects that utilize smart contracts and those intending to commit their cryptocurrencies to smart contracts.
}
Although the actual transactions can reveal transaction patterns and some potential risks, this can not be achieved for smart contracts with few or no actual transactions.
Therefore, analyzing the smart contracts' source code and understanding how they work before engaging in transactions is crucial~\cite{wan2021smart}.

\begin{figure}[!htbp]
\centering
\includegraphics[width=3.45in]{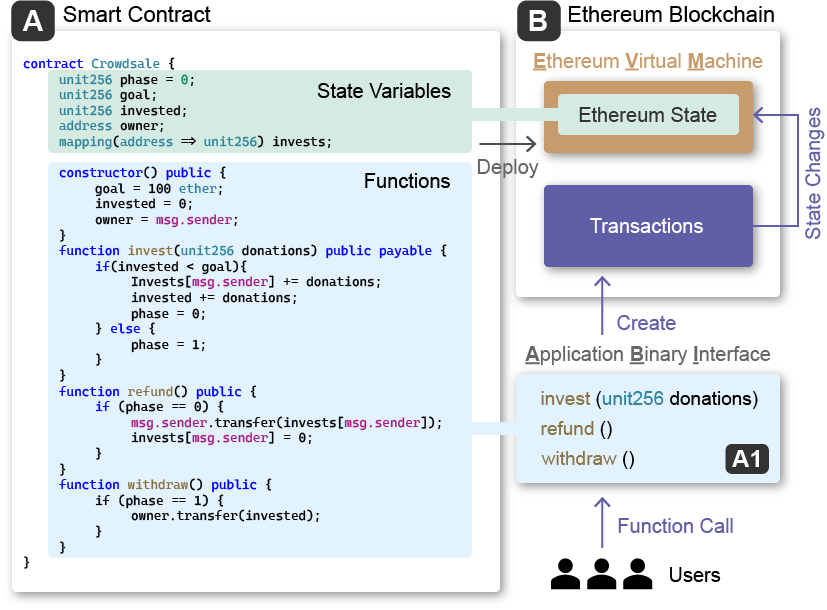}
  \setlength{\abovecaptionskip}{-0.1cm}
  \setlength{\belowcaptionskip}{-0.6cm}
\caption{
Critical concept illustrations: (A) presents an example of the smart contract source code in Solidity, which can be deployed on the Ethereum blockchain (B). Users make a function call via its ABIto execute the contract (A1). If this function call changes the Ethereum state, a transaction will be created to record it on the blockchain.}
\label{fig:background}
\end{figure}

\mod{
However, this is a challenging task for investors, and the challenges differ for investors with and without a programming background.
Most investors lack a programming background needed to understand smart contract functions.
These smart contracts are typically written in programming languages like Solidity~\cite{soud2023dissecting} (Fig.~\ref{fig:background}A), which are difficult for common investors without a programming background to grasp the underlying execution logic.
Even though investors with certain programming knowledge can understand the functions of smart contracts, it is still hard to infer all the potential smart contract behaviors from the source code alone.}
A smart contract's behavior depends on the blockchain's current state, which is influenced by all previous transactions. Real-world scenarios involve multiple investors executing various functions, which can not be easily predicted from the code alone. Additionally, smart contracts may also have unexpected behaviors
due to smart contract defects~\cite{chen2020defining}.
\mod{
Therefore, there is a significant demand for tools that help investors intuitively grasp the functionality and possible behaviors of smart contracts.
}

\mod{To achieve this goal, we face two major challenges. 
\textbf{(C1)} \textbf{How can we get all potential behaviors of a smart contract involving various investors from the source code?}
On the Ethereum blockchain,
smart contract users, including investors,
interact with smart contracts by calling functions via the \textit{Application Binary Interface} (ABI)~\cite{zheng2021application}, 
as shown in Fig.~\ref{fig:background}A1. 
These functions are then executed on \textit{Ethereum Virtual Machine} (EVM), leading to changes in the Ethereum state, internal transactions, and cryptocurrency transfers (e.g., the Ether).
While the behaviors resulting from the function execution can illustrate a smart contract's functionality, they are difficult to observe before actual transactions occur.
Some existing tools like Remix~\cite{Remix} allow users to execute function calls locally to observe this information. However, these tools often require users to manually customize function calls, making it challenging to collect all potential smart contract behaviors involving multiple users with diverse function call patterns.
\textbf{(C2)} \textbf{
How can we intuitively convey the complex behaviors of smart contracts to investors?}
Understanding how a smart contract operates requires analyzing the function calls from multiple users over time, as well as subsequent behaviors of the smart contract, such as cryptocurrency transfers.
These behaviors can exhibit different patterns under different conditions (e.g., function call orders, cryptocurrency value, and state variable values), some of which depend on previous user actions.
Additionally, these behaviors have interrelations; for instance, the recipients and amounts of cryptocurrency transfers may depend on specific state variables, and internal transactions may result in balance changes.
It is challenging to intuitively interpret these complex behaviors to investors, regardless of their technical backgrounds.
}
%
%


To address the aforementioned challenges, we propose \textbf{\techName{}}, a novel visualization approach to enable intuitive visual interpretation of smart contracts without displaying complex source code.
Specifically, we conducted a preliminary study with six domain experts to collect design requirements for understanding smart contracts.
To tackle \textbf{C1}, we developed a smart contract \textbf{simulator} capable of simulating real-world scenarios in which multiple smart contract users invoke various functions.
The simulator collects the necessary information for investors to understand the various behaviors of a smart contract, such as state changes and cryptocurrency flows.
To address \textbf{C2}, we proposed a barcode-based design called \textit{\mO{}}
and an augmented sequential visual design called \textit{\mD{}} within \techName{} to intuitively reveal the smart contract behaviors collected from the simulator. 
Specifically, \textit{\mO{}} is designed to provide a visual summary of involved functions and each address's balance changes in each simulation, and \textit{\mD{}} is designed to reveal the details of each simulation, including the function call sequences, cryptocurrency flows, and state changes.

\mod{
With \techName{}, investors can intuitively understand and explore the behaviors of smart contracts across various scenarios.
To evaluate its effectiveness and usability, we conducted two case studies and in-depth interviews with 12 investors (six with programming backgrounds and six without)}.
The results indicate that \techName{} can help investors understand smart contracts.
Our major contributions are as follows:
\begin{itemize}
    \item We formulate the design requirements for visually interpreting smart contracts, through collaboration with six domain experts.
    
    \item We propose \techName{}, a novel visualization approach that enables intuitive visual interpretation of smart contracts.
    It includes a simulator to capture potential smart contract behaviors for visualization and two visualization modules: a barcode-based design called \textit{\mO{}} to overview the simulations and an augmented sequential design called \textit{\mD{}} to show the details in each simulation.
    
    \item We conduct two case studies and in-depth user interviews with 12 cryptocurrency investors to demonstrate the effectiveness and usability of \techName{}.
\end{itemize}

\section{Background}
In this section, we introduce the background knowledge involved in this study, including smart contracts and smart contract fuzzing.

\subsection{Smart Contracts}
\mod{\textbf{\textit{Smart contracts}} are essentially programs deployed on a blockchain platform and can be executed when predefined conditions are met.}
\textit{Ethereum}~\cite{wood2014ethereum} is one of the most well-known blockchains with smart contracts incorporated.
While smart contracts are also supported by other 
blockchains, we focus on Ethereum as an example in this study. 
The \textbf{\textit{source code}} of smart contracts can be written in various languages (e.g., Solidity~\cite{soliditylang} and Vyper~\cite{vyperlang}), where Solidity is the most popular language for smart contracts on Ethereum~\cite{soud2023dissecting}.
Fig.~\ref{fig:background}A shows the Solidity source code of a smart contract for crowd-sale. 
The source code includes many components like state variables, functions, and events, among which state variables and functions are the two critical parts for understanding the smart contract functionality (green and blue in Fig.~\ref{fig:background}A).
\textbf{\textit{State variables}} are variables whose values are permanently stored in the blockchain, including different types, such as value type (e.g., integers and address), and reference type (e.g., array). 
\textbf{\textit{Functions}} are executable units of code and some of them can be called by external users with their \textit{\textbf{Application Binary Interface (ABI)}}, as shown in Fig.~\ref{fig:background}A1.
In investment-related smart contracts, state variables store investor information or investing conditions while functions indicate investors' potential interactions with contracts, like investing and withdrawing. 
After deploying a smart contract, it can be executed in the \textbf{\textit{Ethereum Virtual Machine (EVM)}}, a stack-based run-time environment of Ethereum (Fig.~\ref{fig:background}B).
The state variables are stored in the \textit{\textbf{Ethereum state}}. The \textit{\textbf{balances}}, cryptocurrencies held by addresses, are also state variables but not defined in the source code.

Both smart contracts and smart contract users have concrete addresses.
When a user makes a \textbf{\textit{function call}} via its ABI, the code of this function will be executed in the EVM.
If this function execution has changed the Ethereum state (e.g., state variables), there will be a \textbf{\textit{transaction}} from the caller's address to the smart contract address created on the blockchain. 
During the function execution, the smart contract will further conduct some transactions (e.g., distributing cryptocurrencies to others) if some triggers or conditions are met.
These transactions from the smart contract address to other addresses, called \textbf{\textit{internal transactions}}, can result in \textit{\textbf{cryptocurrency transferring (flows)}} and change the \textbf{\textit{balance}} of each address respectively. 
Therefore, to understand a smart contract, it is important to figure out the possible behaviors of each function call, such as state variable changes, balance changes, cryptocurrency flows, and internal transactions.

\subsection{Smart Contract Fuzzing} 
\textbf{\textit{Smart contract fuzzing}} has emerged as a powerful technique for finding security bugs in complicated real-world smart contracts, including white-box, black-box, and grey-box fuzzing~\cite{liu2023rethinking}. Among them, \textbf{\textit{coverage-guided grey-box fuzzing (CGF)}} is one of the most successful fuzzing techniques~\cite{wustholz2020harvey}. Its goal is to find test inputs that maximize code coverage within a specified time~\cite{nguyen2020sfuzz}, which is utilized in our simulator to generate comprehensive function call sequences.


\textbf{Fuzzing procedure:} The inputs of the CGF fuzzer (the tool conducting smart contract fuzzing) include the smart contract to be tested, the set of original user-provided test inputs (function calls), and the expected fuzzing time. Initially, the fuzzer generates new test inputs by mutating a set of existing ones, called \textit{seed pool}, which initially contains only the user-provided test inputs. 
During each iteration, the fuzzer randomly picks a test input from the seed pool, mutates it to generate new test inputs, and executes the mutated inputs with the target contract. 
The execution is monitored by the fuzzer to collect the necessary information. If the test input execution triggers buggy behaviors (e.g., reentrancy~\cite{liu2018reguard}),
the test input is added to a \textit{bug pool}. 
If the test input executes new branches in the contract, it is added to the seed pool and will be further mutated to generate more test inputs in the next iteration. 
Test inputs that do not cover new branches are considered irrelevant and are discarded. 
Once the time budget is exhausted, all test inputs in the seed pool and the bug pool are returned.
Since the seed pool can cover most unique branches in the source code, we use it to generate the simulated function call sequence in our simulator.
\section{Related Work}
This work is related to prior research on \textit{algorithm-based smart contract analysis}, \textit{smart contract visualization}, and \textit{software visualization}.

\subsection{Algorithm-based Smart Contract Analysis}
Recently, much research on algorithm-based smart contract analysis has been proposed to detect security vulnerabilities~\cite{liu2019survey}, which can be mainly divided into static and dynamic analysis~\cite{kushwaha2022ethereum}.
\textbf{Static analysis} relies on code analysis without execution to detect issues in the smart contract~\cite{feist2019slither}, and encompasses various approaches, such as taint analysis, symbolic execution, formal verification, and machine learning~\cite{qian2023mufuzz}.
Taint analysis detects whether the untrusted information will spread to critical program points~\cite{brent2020ethainter,ghaleb2022etainter,liang2024ponzicuard}. 
Symbolic execution collects potential execution paths of smart contracts and then uses Z3 solver~\cite{de2008z3} to emulate their execution~\cite{luu2016making,krupp2018teether,chen2021sadponzi}.
Formal verification uses mathematical methods to prove the code correctness~\cite{bhargavan2016formal,abdellatif2018formal,murray2019survey,garfatta2021survey}. 
Machine learning detects vulnerabilities based on labeled smart contracts and customized models~\cite{tann2018towards,wang2020contractward,liu2021combining}.
\textbf{Dynamic analysis} identifies security issues in smart contracts by directly running the code, often through fuzzing tests~\cite{qian2023mufuzz}.
\textit{ContractFuzzer}~\cite{jiang2018contractfuzzer} firstly utilizes fuzzing to detect vulnerabilities by observing runtime behavior. 
Subsequent research has focused on enhancing aspects of smart contract fuzzing, including feedback-driven fuzzers like \textit{ContraMaster}~\cite{wang2020oracle}, \textit{Harvey}~\cite{wustholz2020harvey}, and \textit{Echidna}~\cite{grieco2020echidna}, which generate varied inputs to uncover bugs effectively. 
\textit{sFuzz}~\cite{nguyen2020sfuzz} introduced branch distance feedback for exploring difficult branches, while Liu et al.~\cite{liu2023rethinking} developed a strategy for generating sequences that activate deeper contract states to detect complex vulnerabilities.
Qian et al.~\cite{qian2023mufuzz} introduced sequence-aware mutation and seed mask guidance to achieve high branch coverage and bug finding.

The above studies focus on ensuring the correctness of the code execution, without considering aiding in understanding the smart contract. 
In this paper, we developed a simulator based on \textit{smart contract fuzzing} (one of the above dynamic approaches) and proposed visualizations based on the simulator to help investors understand smart contracts.

\subsection{Smart Contract Visualization}
Due to the complexity of blockchain technology, various visualizations have been proposed to help understand the blockchain intuitively~\cite{sundara2017study,tovanich2019systematic,tovanich2019visualization}, but only a few studies focus on smart contract visualizations, which can be divided into transaction-based and code-based approaches based on the visualized data type~\cite{harer2019comparison}.
\textbf{Transaction-based approaches} focus on visualizing blockchain transactions associated with smart contracts and are mainly used for analyzing the transaction patterns~\cite{yap2023smart,sourekirin,yan2023nftvis,cao2023nfteller} or detecting frauds that have already occurred~\cite{jeyakumar2023visualizing,wen2023nftdisk}.
\mod{
These approaches are designed to highlight the network structure of cryptocurrency flows, without investigating smart contracts' functions (e.g., how these flows are produced).
They also do not work well for smart contracts with few or no transactions.
}
\textbf{Code-based approaches} focus on displaying the logic and structure within smart contract code to aid in understanding smart contract functionality~\cite{bragagnolo2017smartinspect,skotnica2019contract,pierro2021smart}. 
Furthermore, some studies visualize the control flow graphs of smart contracts to track the actual execution process~\cite{norvill2018visual} and identify vulnerabilities before deploying~\cite{wen2023code,wen2024ponzilens+}.
\mod{
However, these approaches are not intuitive for investors and can not reflect real-world scenarios involving various function calls from different users.
}

\mod{
Unlike previous approaches, we take the smart contract's source code as the input
and offer an intuitive interpretation of smart contract functionality by visualizing both simulated function calls and related smart contract behaviors like transactions.
}

\subsection{Software Visualization}
Previous software visualizations mainly focus on three aspects of software: structure, evolution, and behavior~\cite{diehl2007software,chotisarn2020systematic}.
Structure visualization aims to reveal the source code hierarchies~\cite{hayatpur2023crosscode}, package dependencies~\cite{isaacs2018preserving}, and execution logic~\cite{devkota2021cfgconf}.
Evolution visualizations analyze the history of code changes to illustrate the development and progression of a software system~\cite{yoon2013visualization,kim2020githru}.
Behavior visualizations present data from software execution, such as time series and domain-specific event sequences, to support performance analysis and anomaly detection~\cite{zhou2024fctree,xu2019clouddet,guo2018valse}.
Our approach falls under software behavior visualization, as we focus on visualizing the potential behaviors of smart contracts. However, rather than concentrating on execution details, we visualize the overall behaviors of smart contracts when they interact with simulated user behaviors. 
Compared with previous work, our visual designs support the analysis of complex scenarios specific to smart contracts by integrating users' function calls with cryptocurrency flows, balances, and state changes.
\section{Informing the Design}
To 
inform our subsequent
visualization designs, 
we conducted a preliminary study with six domain experts to collect design requirements. 
This section reports the feedback distilled from the preliminary study.

\subsection{Preliminary Study}\label{Sec:preliminary}
The preliminary study aims to collect the requirements for designing visualizations to help investors understand smart contracts without having to deal with complex source code.
We conducted the study under the guidance of design study methodology in Sedlmair et al.'s work~\cite{sedlmair2012design}.
The participants and procedures are as follows.

\textbf{Participants:} 
The study involved six domain experts \textbf{(E1-E6)} with extensive experience in smart contract investments.
\textbf{E1-E3} can analyze smart contract source code due to their daily work experience.
\mod{\textbf{E1 and E2}} work in a blockchain security company, with \textbf{E1} being a smart contract auditor and \textbf{E2} a programmer.
\textbf{E3} is a Post-doc whose research focuses on smart contract security.
\mod{\textbf{E4-E6} are investors with over three years of experience in investing in smart contract applications, but they do not have a background in analyzing smart contract source code.}

\textbf{Procedures:}
\mod{
The study has two sessions.
In the first session, we conducted one-on-one, semi-structured, hour-long interviews with each expert (\textbf{E1-E6}).
Experts were asked to describe the challenges they face in understanding smart contracts and outline their expectations for visual interpretations. Their feedback was textually recorded.
By summarizing their feedback, we then derive the initial design requirements, including simulating real-world usage scenarios with multiple users, highlighting cryptocurrency flows and balance changes, and incorporating necessary information like state variables.
In the second session, we presented these initial design requirements to each expert and asked them to review the proposed requirements in the context of their experiences.
We also provided a draft visual design for reference.
Feedback from this session included suggestions for modifying requirements and comments on the draft design (e.g., providing summaries of each function's behaviors and highlighting state variable changes).
This feedback was used to finalize the design requirements and ensure that the visualizations could meet the experts' expectations.
}

\subsection{Design Requirements}
We summarized six design requirements from the preliminary study:

\begin{itemize}
    \item[\textbf{R1}] \textbf{Collect possible behaviors of a smart contract via simulation.} All experts (\textbf{E1-E6}) mentioned that our tool should demonstrate the most possible behaviors of a smart contract to help investors understand the smart contract comprehensively.
    Since the source code is the only input of our tool and the behaviors of smart contract users are uncontrollable in the real world, all experts (\textbf{E1-E6}) suggested that a simulation of scenarios where multiple smart contract users invoke various functions is helpful for collecting the possible behaviors of smart contracts.
    Three experts (\textbf{E1-E3}) suggested that the simulation can refer to smart contract fuzzing, a technique for vulnerability detection, which requires covering as many branches in the source code as possible to ensure a comprehensive exploration of smart contract behaviors~\cite{nguyen2020sfuzz}. 
    
    \item[\textbf{R2}] \textbf{Provide an overview of each simulation.} Three experts (\textbf{E1-E3}) emphasized that the simulation could be conducted multiple times to cover various scenarios, so our visualizations have to support the overview of each simulation. \textbf{E4} and \textbf{E6} also suggested that it is necessary to provide a summary of addresses' earnings and losses for a deep understanding of smart contracts. 
    \textbf{E2} mentioned that showing the functions involved can enrich overview.
     
    \item[\textbf{R3}] \textbf{Show the function call sequences.}
    All experts (\textbf{E1-E6}) suggested that our visualization should demonstrate the pattern of the simulated function call sequence, including callers, function names, and the order of the function call sequence. They emphasized that showing the function call behaviors of each address is the premise of understanding the function call distribution and deducing the functionalities of smart contracts.
    \item[\textbf{R4}] \textbf{Show the cryptocurrency flows.} Three experts (\textbf{E4-E6}) mentioned that it is critical to display the cryptocurrency flows caused by the internal transactions of smart contracts after each function call. This feature is helpful for investors without a programming background to understand the smart contract intuitively. \mod{\textbf{E1 and E2}} pointed out that cryptocurrency flows are also important for investors to verify the economic model of the smart contract, which can hardly be achieved by checking the source code.
    
    \item[\textbf{R5}] \textbf{Show the state variable changes.}
    \textbf{E1-E3} suggested that it is necessary to show the state variable changes after each function call since not all functions will invoke internal transactions, and state variable changes can help understand such functions. Three experts \mod{\textbf{(E1, E4, E5)}} mentioned that the relations between state variable changes and cryptocurrency flows can also help investors deduce the intrinsic logic of each function.
    \item[\textbf{R6}] \textbf{Show the net balance changes.} All experts (\textbf{E1-E6}) agreed that the changing balance of both the smart contract and smart contract users indicate their earnings or losses caused by the users' behaviors (i.e., the sequence of function calls), which is one of the most intuitive pieces of evidence for analyzing the functionality of smart contracts.
    \textbf{E4} added that analyzing the net balance changes over time is useful for identifying the type of smart contracts and revealing the potential risks for investors. 
\end{itemize}

\section{Simulator}
\begin{figure}
  \setlength{\abovecaptionskip}{0.1cm}
  \setlength{\belowcaptionskip}{-2em}
    \centering
    \includegraphics[width=1\linewidth]{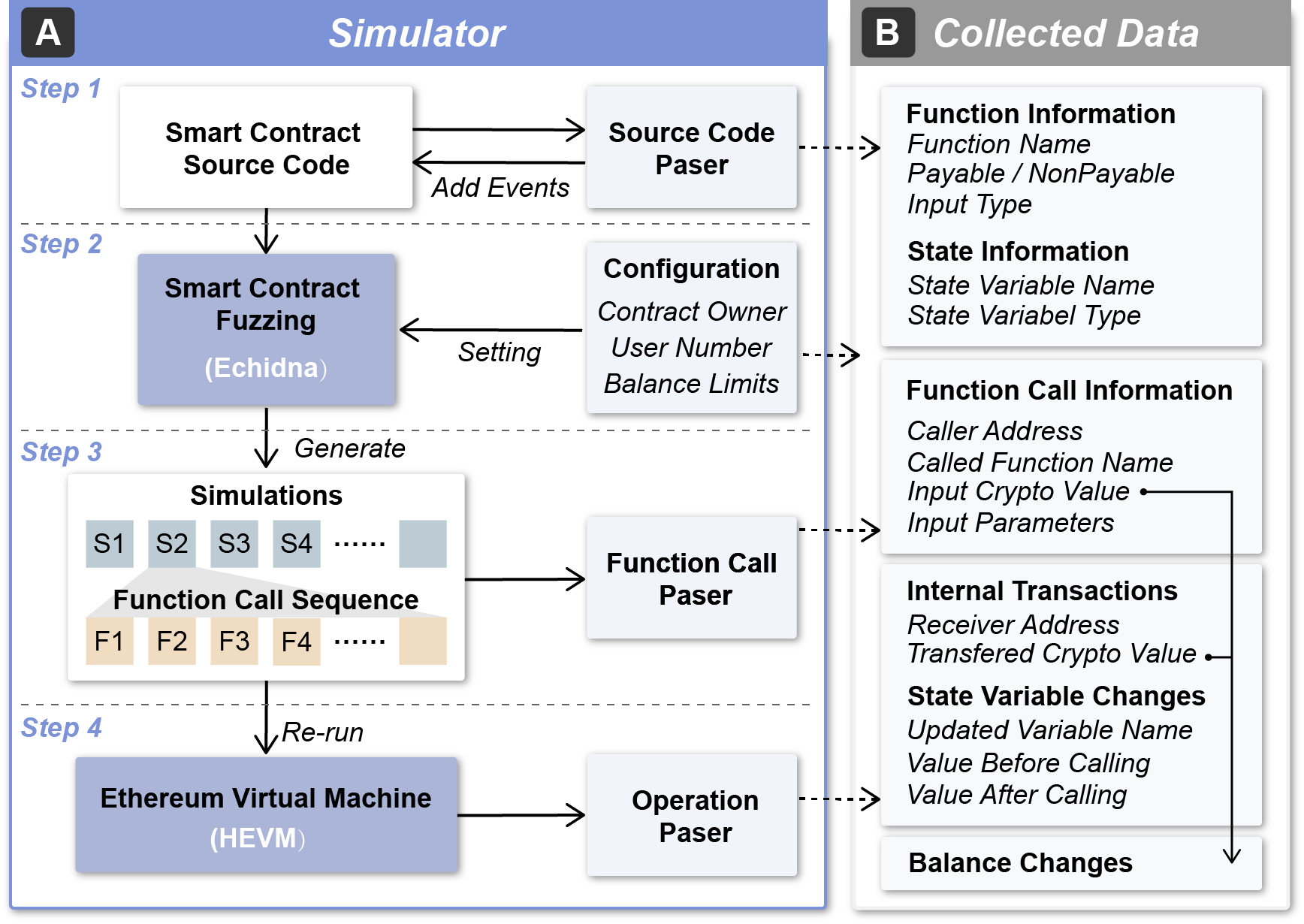}
    \caption{The simulator framework (A) consists of four steps: source code parsing, fuzzing configuration, function call parsing, and operation parsing. (B) shows the collected data for the visualizations.}
    \label{fig:simulator}
\end{figure}

To collect necessary data for visualizations, we developed a simulator to simulate all the possible behaviors of a smart contract in real-world scenarios, where multiple users call various functions (\textbf{C1}).
The collected data, such as state variable changes and internal transactions, is further visualized by our visualization approach (Section~\ref{sec-visualization}) to enable an intuitive visual interpretation of smart contracts. 
We incorporated a smart contract fuzzing tool to generate function call sequences involving various users and functions, as it can cover most unique branches in the source code (\textbf{R1}).
Fig~\ref{fig:simulator} illustrates the overall architecture of our simulator framework (Fig~\ref{fig:simulator}A) and provides a detailed view of the collected data (Fig~\ref{fig:simulator}B). The framework can be divided into four steps:

\textbf{Step 1: Source Code Parsing.}
Given that the source code of a smart contract is the input for our simulator, we develop a source code parser to identify information about state variables and functions, such as their names and types.
This is achieved by parsing the \textit{Abstract Syntax Tree (AST)} of the source code, which can be generated by the \textit{Solidity Compiler}~\cite{soliditylang}.
Additionally, the parser inserts events before and after each function in the contract source code to capture the specific values of state variables before and after each simulated function call.

\textbf{Step 2: Fuzzing Configuration.}
We input the source code into an existing smart contract fuzzing tool, \textit{Echidna}~\cite{grieco2020echidna}, which is an industry-standard tool widely used in many popular projects (e.g., 0x~\cite{0xprotocol} and Balancer~\cite{balancercore}).
We provide a default configuration of the fuzzing tool, and users can modify it to fit their requirements, such as setting user numbers, defining balance limits for addresses, and specifying the contract owner. These configurations are also used in visualizations.

\textbf{Step 3: Function Call Parsing.}
\mod{
Based on the configurations and source code, the fuzzing tool generates multiple simulations to cover as many source code branches as possible.
The number of simulations depends on the tool's capabilities, smart contract code complexity, and settings specified in Step 2, such as the number of simulated users.
All state variables and functions in the source code are considered.
}
Each simulation contains a function call sequence involving multiple users and functions. We develop a function call parser to collect detailed information for each function call in the sequence, including the function name, caller address, input cryptocurrency value, and input parameters.

\begin{figure*}
    \centering
      \setlength{\abovecaptionskip}{0.1cm}
  \setlength{\belowcaptionskip}{-2em}
    \includegraphics[width=0.9\linewidth]{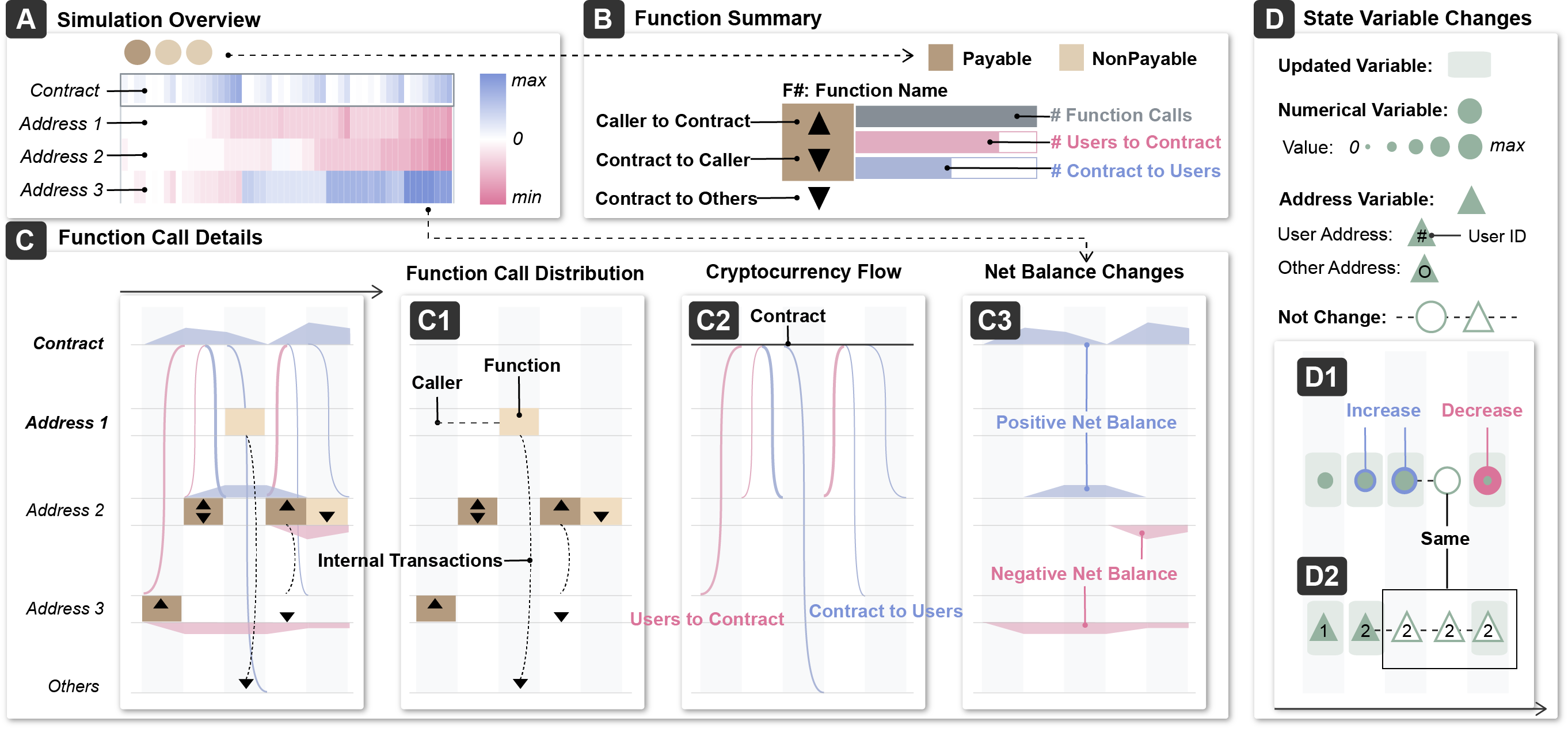}
    \caption{The visual designs of \techName{}. The \mO{} (A) shows a visual summary of each simulation. The \mD{} includes: Function Summary (B) to overview each function, Function Call Details (C) to show the function call distribution (C1), cryptocurrency flow (C2), and net balance changes (C3), and State Variable Changes (D) to show the variables changed in each function call.  }
    \label{fig:design}
\end{figure*}

\textbf{Step 4: Operation Parsing.}
Subsequently, we rerun the simulated function call sequence on \textit{HEVM}, a stand-alone Ethereum Virtual Machine, to capture the execution traces.
An operation parser is built to analyze the specific operations within the execution traces, where we extract internal transactions from the operations with operation codes like ``\textit{CALL}" and ``\textit{CALLCODE}", and gather variable changes from the operation code ``\textit{LOG}", triggered by the events inserted in Step 1.

\section{\techName{}}\label{sec-visualization}
Built upon the data collected from our simulator, we propose \techName{}, a novel visualization approach to provide cryptocurrency investors with an intuitive visual interpretation of smart contracts.
Fig.~\ref{fig:teaser} displays the interface of \techName{}, including two modules: \textit{\mO{}} (A) and \textit{\mD{}} (B).
The \mO{} shows a visual summary of each simulation from the simulator.
The \mD{} visualizes the details of each simulation, including \textit{Function Summary} (Fig.~\ref{fig:teaser}B1), \textit{Function Call Details} (Fig.~\ref{fig:teaser}B2), and \textit{State variable Changes} (Fig.~\ref{fig:teaser}B3).
With \techName{}, investors can first observe each address's balance changes in the \mO{} and delve into the \mD{} of each simulation by clicking it to check the detailed behaviors of the smart contract after each function call.


\subsection{\mO{}}
The \mO{} (Fig.~\ref{fig:design}A) is a barcode-based design that provides a visual summary for each simulation, involving the net balances over time and the involved functions (\textbf{R2}).
Specifically, the Y-axis lists the addresses of the contract and simulated users, and the X-axis indicates the function call sequence in a relative time order.
The color of each cell encodes the net balance of each address \mod{(including the contract address)} after each function call, with a gradient color scheme from red~\red\, to blue ~\blue\, indicating the value from negative to positive.
With this design, investors can easily understand the earnings and losses of every address after each function call in the function call sequence.
\mod{Additionally, we use circles at the top to show the functions called in this simulation}, where the dark color~\dark\, encodes the payable function while the light color~\light\, encodes the non-payable function.
The payable function means users can send cryptocurrency to the contract with this function, while the non-payable function can not carry any cryptocurrency.
\mod{
Investors can overview how many unique functions are involved in each simulation and their types by checking these circles.
}

\subsection{\mD{}}
The \mD{} is an augmented sequential visual design that demonstrates detailed information in each simulation, including three parts: Function Summary (Fig.~\ref{fig:design}B), Function Call Details (Fig.~\ref{fig:design}C), and State Variable Changes (Fig.~\ref{fig:design}D).

\textbf{Function Summary} shows the statistics of the calls of each function in a simulation.
Specifically, the visual encoding of each function is shown in Fig.~\ref{fig:design}B.
We use a rectangle to indicate each function, and the color of this rectangle is the same as the \mO{} for payable~\dark\, or non-payable~\light.
We also show the function name above the rectangle.
Three types of triangles are used to demonstrate how the function transfers the cryptocurrencies.
An upward triangle inside the rectangle~\upin\, means that users have called this function to send cryptocurrencies to the contract in this simulation, indicating that there is at least a cryptocurrency flow from the caller (i.e., who called this function) to the contract.
A downward triangle inside the rectangle~\downin\, indicates that this function has been used to \mod{trigger} the contract to send cryptocurrency to the caller itself.
A downward triangle out of the rectangle~\downout\, means that this function has been used to \mod{trigger} the contract to transfer cryptocurrency to addresses other than the caller.
There are three horizontal bars on the right of the rectangle designed to show the number of function calls of different cryptocurrency transferring patterns.
The length of the first grey bar~\lgrey\, indicates the number of function calls of this function.
The length of the second red bar~\lred\, shows the number of calls that send cryptocurrencies to the contract in these function calls, while the third blue bar~\lblue\, shows the number of calls that \mod{trigger} the contract to send cryptocurrency to user addresses.

\textbf{Function Call Details} (Fig.~\ref{fig:design}C) demonstrates the detailed behaviors of the smart contract after each function call, which can be regarded as a combination of three layers: the \textit{function call distribution} layer (Fig.~\ref{fig:design}C1), the \textit{cryptocurrency flow} layer (Fig.~\ref{fig:design}C2), and \textit{net balance changes} layer (Fig.~\ref{fig:design}C3).
Layout of Function Call Details is aligned with the \mO{}, i.e., the X-axis for the function call sequence and the Y-axis for the addresses, where we highlight the owner of the contract in bold font and add an ``\textit{others}" label at the bottom to involve the cryptocurrency transfers with addresses other than the simulated users.  
Additionally, a white-gray striped background is used to help distinguish each function call in the sequence.

The \textit{\textbf{function call distribution}} layer (Fig.~\ref{fig:design}C1) shows the simulated function call sequence among multiple users (\textbf{R3}), where the function calls from each address are encoded by the rectangle on the row of the respective address, with the rectangle color the same as the Function Summary. 
Similar to the Function Summary, we use black triangles to highlight the feature of cryptocurrency transfer in this function call.
Inside a column of a function call, the upward~\uptri\, and downward~\downtri\, triangles at the row of an address indicates a transfer from this address to the contract or from the contract to this address, respectively, after this function call.
As shown in Fig.~\ref{fig:design}C1, we use a dashed line~\dline\, to encode the internal transactions of a function call, where the dashed line links the caller calling this function and the receiver of the internal transactions, indicating that the caller \mod{triggers} the contract to send cryptocurrencies to the receiver. 
With this layer, investors can identify the function call distribution and the pattern of internal transactions.

The \textit{\textbf{cryptocurrency flow}} layer (Fig.~\ref{fig:design}C2) shows how the cryptocurrency is transferred in detail (\textbf{R4}).
Specifically, a red curve~\fout\, from the row of one user address to the contract indicates that the user pays the cryptocurrencies to the contract in this function call, while the blue curve~\fin\, from the contract to the row of a user address means that the contract sends cryptocurrencies to this user by conducting an internal transaction.
The width of curves encodes the value of cryptocurrencies within the cryptocurrency flows.
With this layer, investors are allowed to analyze the concrete cryptocurrency flows in each function call.

The \textit{\textbf{net balance changes}} layer uses area charts to visualize the net balance changes of each address (\textbf{R6}), which are calculated based on the cryptocurrency flows. 
The blue~\lblue\, areas above the row of addresses and the red~\lred\, areas below encode the positive and negative net balances of each address, respectively, where the height of these areas indicates the concrete value of cryptocurrency balances.
With this layer, investors can understand the patterns of each address' earnings or losses.

\textbf{State Variable Changes.}
As shown in Fig.~\ref{fig:teaser}B3, we visualize the state variable changes (\textbf{R5}) at the bottom of the \mD{} to show how the state variables are changing after each function call.
We only show the variables of the value type, such as integers and addresses, since the reference types, like mappings and arrays, have complex intrinsic structures which make it hard to determine the relations between them and the functions.
As shown in Fig.~\ref{fig:design}D, the State Variable Changes share the same X-axis with Function Call Details to indicate the function call sequence for an easy observation of the relations between function calls and state variables changed by them.
The Y-axis lists the state variables that have been changed in this simulation.
On the left, we show their variable name and type defined in smart contracts and icons indicating their type.
We use green circles~\gcircle\, for numerical variables and green triangles~\gtri\, for non-numerical variables, as shown in Fig.~\ref{fig:design}D.
For the state variable changes in each function call, we first fill a light green background~\lgreen\, at the crossing of the related variable and function call, indicating this variable has been updated in this function call.
Then, we visualize each variable's concrete value and changes of each function call by showing an icon in the same shape as mentioned above.
Specifically, we use the radius of the circle~\circleR\,  to show the value of numerical variables and add a red~\rborder\, or blue~\bborder\, border around the circle, with the border width indicating the decreased or increased value in this function call, as shown in Fig.~\ref{fig:design}D1.
For address type, we add a text label on the triangle~\addr\, to show their identity, i.e., a number for the address index of a simulated user and ``O" for addresses~\Oaddr\, other than them, as shown in Fig.~\ref{fig:design}D2.
For the variables that have not been changed, we fill them in white~\fcircle~\ftri\, and use a horizontal dashed line~\hdline\, to link them with the previous one to indicate that no changes happened to them.

\subsection{Interactions}
\begin{figure*}
    \centering
      \setlength{\abovecaptionskip}{0.1cm}
  \setlength{\belowcaptionskip}{-0.6cm}
    \includegraphics[width=0.92\linewidth]{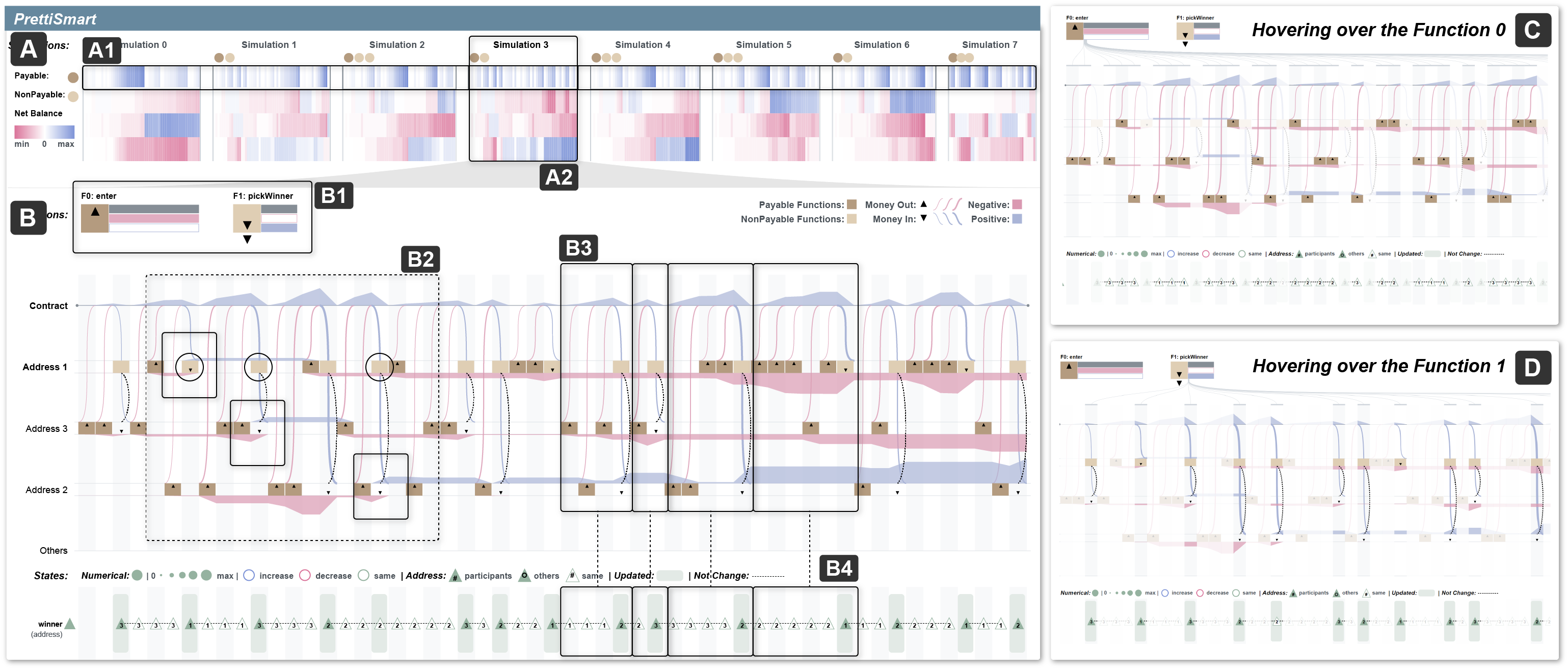}
    \caption{With \techName{}, an investor has identified a smart contract as a fair gambling game. (A) shows the overview of simulations from our simulator, where the patterns in (A1) and (A2) help the investor understand the gains and losses of each address. (B) helps analyze the function summaries (B1), the reasons for turning losses into wins (B2), and the repetitive patterns in cryptocurrency flow (B3) and state variable changes (B4). The user checked the two involved functions in (C) and (D) by hovering over the function summaries.}
    \label{fig:case1}
\end{figure*}

\mod{
\techName{} enables rich interactions to allow investors to smoothly explore smart contract behaviors.
In the \mO{}, investors can click one simulation to delve into the simulation details.
In the \mD{}, investors can hover over the function summary to highlight information related to this function in Function Call Details and State Variable Changes, \mod{as shown in Fig.~\ref{fig:case1}D.}
We also allow investors to click on function rectangles and variable icons to view underlying data through pop-up tooltips, including total function call counts, inputs of each function call, and specific variable values. 
}

\section{Case Study}\label{sec:case}
\mod{
This section presents two case studies to demonstrate the effectiveness of \techName{}.
They were conducted by two (U8 and U1) of the twelve participants (U1-U12) in our user interviews (see Section~\ref{sec:interview}).
}

\subsection{Case 1: Interpreting a Gambling Contract}
U8 is an investor with three years of investment experience in various smart contract applications but lacks a background in reading the smart contract source code. 
During the user interview, he was asked to analyze a smart contract with \techName{}.
Upon launching \techName{}, he first explored the \mO{} to observe the net balance changes of each simulated user and the contract, as shown in Fig.~\ref{fig:case1}A.
In Fig.~\ref{fig:case1}A1, he noted that the contract's balance patterns were repetitive across simulations, i.e., the blue~\blue\, gradually darkens, then abruptly shifts to white, indicating a gradual increase in the contract's balance followed by a sudden drop to zero.
U8 found that all had experienced both gains (blue~\blue\,) and losses (red~\red\,) across the seven simulations, suggesting that everyone had the possibility to make earnings.

\textbf{Turning losses into wins.}
Since U8 found that all the three simulated users had both earns and losses in \mod{\textit{Simulation 3}}, he delved into its \mD{} (Fig~\ref{fig:case1}B) by clicking on it to analyze the reasons for turning losses into wins. 
In Fig.~\ref{fig:case1}B1, he saw two function summaries \mod{(\textit{F0: enter} and \textit{F1: pickWinner})} involved in this simulation, which may be the main way for users to interact with this contract.
\mod{\textit{F0}} was encoded in a dark color with an upward triangle~\darkup\, inside and a full red bar~\lred\, on the right, indicating it is a payable function used to pay cryptocurrency into the contract.
\mod{\textit{F1}} was colored in a light color with both inside and outside downward triangles~\lightdowndown\, and a full blue~\lblue\, bar, meaning that it is a non-payable function and is used to \mod{trigger} the contract to send cryptocurrencies to both the caller and others.
When he checked the Function Call Details, he noticed three areas indicating the user turned the losses into wins, where the area charts varied from the downside red~\lred\, area to the upside blue~\lblue\, area (squares of Fig.~\ref{fig:case1}B2).
As shown in the circles of Fig.~\ref{fig:case1}B2, he found that all function calls of these three areas were encoded in light rectangles~\light\, at the row of \textit{address 1}, indicating that all the three function calls were made by \textit{address 1} and the called function was \mod{\textit{F1}}, the only one in a light color~\light.
According to the blue curves for cryptocurrency flows, U8 confirmed that the reason for turning losses into wins was that \textit{address 1}, the owner of this contract, called the function \mod{\textit{F1}} to let the contract transfer cryptocurrencies to the user turning losses into wins.

\textbf{Repetitive cryptocurrency flows.}
U8 found repetitive cryptocurrency flows in Fig.~\ref{fig:case1}B3, i.e., multiple red curves~\fout\, with a following blue curve~\fin, indicating the cryptocurrencies were always transferred from user addresses to the contract first and then followed by a transfer from the contract to a certain user address.
The area chart showed that the contract's balance increased gradually and decreased to zero in each repetitive round, indicating the contract sent all cryptocurrencies received before. 
U8 also found some repetitive state variable changes in the same function calls, as shown in Fig.~\ref{fig:case1}B4.
He thought that the behaviors of functions could help him understand these repetitions, so he checked the behaviors of each single function by hovering over it, as shown in Fig.~\ref{fig:case1}C and Fig.~\ref{fig:case1}D.
Then, he found that all function calls of \textit{F0: enter} were used to send cryptocurrencies to the contract, which makes the contract's balance increase gradually.
The only address variable called ``winner" is filled in white~\ftri, indicating they were not changed.
In Fig.~\ref{fig:case1}D, he found that all calls of \textit{F1: pickWinner} had the same caller \textit{address 1} who is the owner of the contract (highlighted in bold font).
This function \mod{triggered} the contract to send cryptocurrencies to simulated user addresses (the blue curves and dashed lines), where the value is exactly the total value of previous investments (the area chart's height changed into zero in Fig.~\ref{fig:case1}D).
By checking the State Variable Changes, he saw that the address variable called \textit{winner} has been changed into the address that received cryptocurrencies in each call of \textit{F1}, suggesting that it may be the variable storing the receiver.
According to these findings, U8 concluded that \textit{F0: enter} is a function for investing cryptocurrencies into the contract, and \textit{F1: pickWinner} is a function that only can be called by the owner and used to select a winner from previous investors to accept all investments.
He guessed that the contract was probably a gambling game, and the repetitive cryptocurrency flows were caused by the game rounds.

In fact, this contract is exactly a gambling game called \textit{lottery}~\cite{lottery}.
\mod{
With \techName{}, U8 tended to believe that this game is likely to be fair since}
he found that the winner was selected randomly by \textit{pickWinner} and the contract sent all its balance to the winner without charging fees.

\begin{figure*}
    \centering
      \setlength{\abovecaptionskip}{0.1cm}
  \setlength{\belowcaptionskip}{-0.4cm}
    \includegraphics[width=0.9\linewidth]{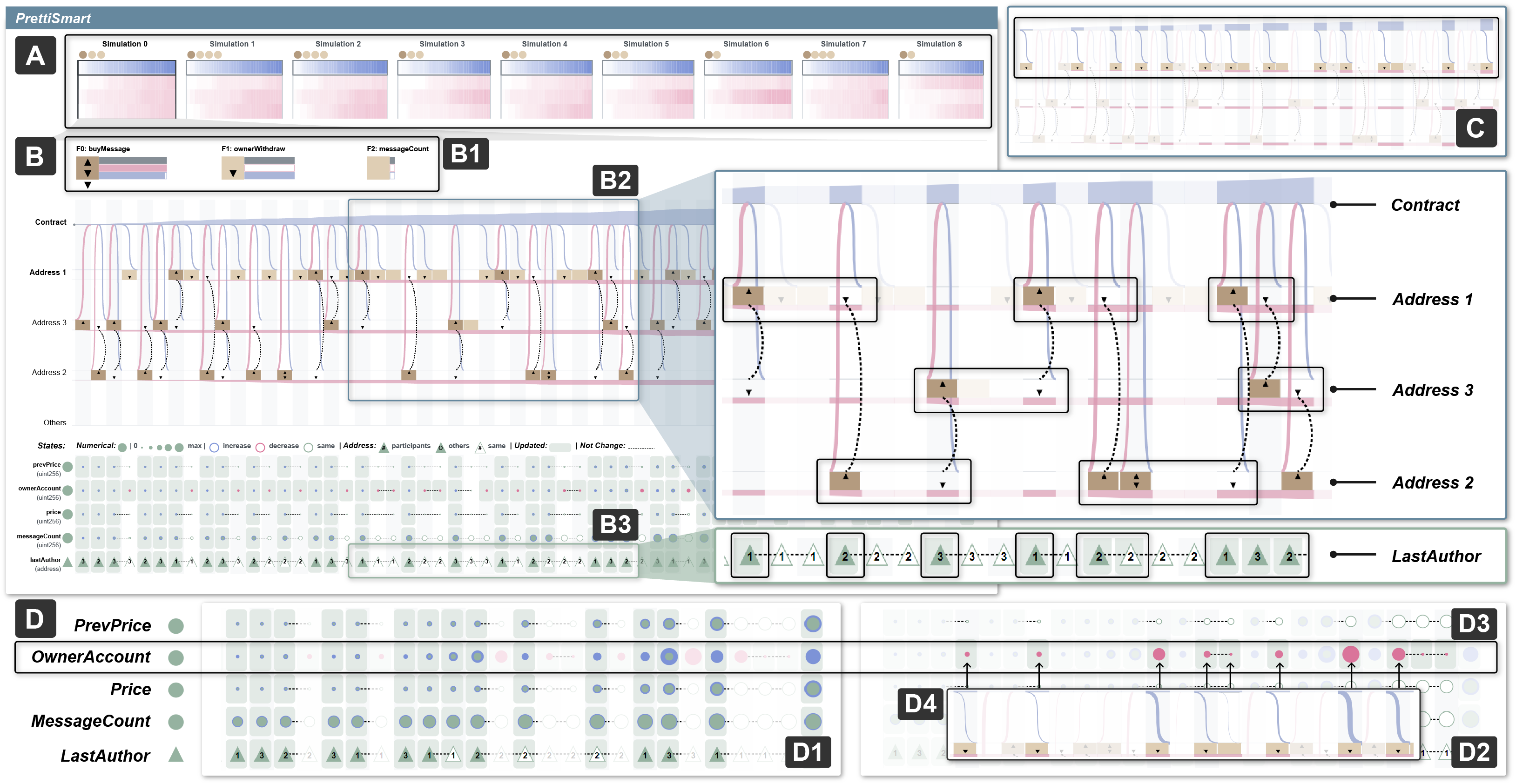}
    \caption{With \techName{}, an investor identified a fraudulent smart contract. (A) shows that all simulated addresses lost their cryptocurrencies and the contract had an increasing balance. By observing the \mD{} (B), the investor found an abnormal chain-like cryptocurrency flow (B2) and a repetitive state change pattern (B3). After checking the function call details of \textit{F0} (C and D4) and comparing the state change pattern of \textit{F0} and \textit{F1} (D1-D3), the investor confirmed that this smart contract is a fraudulent contract.}
    \label{fig:case2} 
\end{figure*}
\subsection{Case 2: Identifying a Fraudulent Smart Contract}\label{sec:case2}
U1 is a PhD candidate with rich experience in smart contract programming.
Initially, U1 browsed the \mO{} (Fig.~\ref{fig:case2}A) and observed that all simulations looked similar in terms of the net balance changes.
Specifically, the contract's balance increased (a gradually darker blue~\blue) and all user addresses' balances were decreasing (A gradually darker red~\red).
This indicated that all simulated users had invested in the contract without receiving a reasonable profit in return, which appeared abnormal and fraudulent for U1.
To find out the reason, he analyzed the \mD{} (Fig.~\ref{fig:case2}B), where the Function Summaries (Fig.~\ref{fig:case2}B1) showed three involved functions.
\textit{F0: BuyMessage} was called most times (the longest grey~\lgrey\, bar among three functions) and encoded in a dark rectangle with all three kinds of triangles~\darkupdowndown, indicating that it has been used to transfer cryptocurrencies both from the caller to the contract and from the contract to any simulated users including the caller. 
\textit{F1: ownerWithdraw} was encoded in a light color with an inside downward triangle~\lightdown, meaning that it was only used to \mod{trigger} the contract to send cryptocurrencies to the caller back. 
\textit{F2: messageCount} was called rarely (the shortest grey~\lgrey\, bar) without any cryptocurrency flows (the empty red~\lred\, and blue~\lblue\, bars).

\textbf{Chain-like internal transactions.}
In the Function Call Details, U1 found it interesting that the dashed lines~\dline\, always linked the current caller and last caller of function \textit{F0} (the only dark one~\dark\, in this simulation).
These dashed lines looked like a chain-like shape, which became more apparent after hovering over \textit{F0}, as shown in Fig.~\ref{fig:case2}B2.
Since the dashed lines were used to link the caller of the function call and the receiver of internal transactions caused by this function call, this chain-like pattern of internal transactions indicated that once the function \textit{F0: BuyMessage} is called the contract will send cryptocurrency to its last caller.
In the State Variable Changes (Fig~\ref{fig:case2}B3), U1 noticed that an address variable named ``\textit{LastAuthor}" changed only in function calls of \textit{F0}, and this variable exactly stored the address of the current caller.
Additionally, there were always two curves in calls of \textit{F0}, i.e., a red curve~\fout\, from the caller to the contract, indicating an investment, and a blue curve~\fin\, from the contract to the address in the \textit{LastAuthor}, indicating a payment to \textit{LastAuthor}.
Based on the above observation, U1 concluded that investors can invest in the contract by calling \textit{F0}, and then the contract directly transfers a portion of the investment (the blue curve~\fin\, is finer than the red one~\fout\,) to the last investor, whose address is exactly stored in the ``\textit{LastAuthor}" variable.

\textbf{Owner ``steals" your investment.}
Next, U1 was curious about \textit{F1: ownerWithdraw}, so he hovered over it to see the Function Call Details (Fig.~\ref{fig:case2}C).
He noticed that all rectangles~\light\, for \textit{F1} were aligned with the row of \textit{address 1}, the owner of this smart contract, suggesting that \textit{F1} probably can only be called by the owner.
Each call of \textit{F1} involved a blue curve~\fin\, from the contract to the owner, leading U1 to suspect that \textit{F1} enabled the owner to withdraw cryptocurrency from the contract.
This revealed a significant risk for investors, as it essentially permitted the owner to ``steal" the investment.
To further assess this risk, U1 analyzed the State Variable Changes of \textit{F0} (Fig.~\ref{fig:case2}D1) and \textit{F1} (Fig.~\ref{fig:case2}D1).
He observed that all state variables were updated (with a light green background~\lgreen\,) in calls of \textit{F0} (Fig.~\ref{fig:case2}D1), while only a numerical variable named ``\textit{OwnerAccount}" was updated (with a light green background~\lgreen\,) by \textit{F1} (Fig.~\ref{fig:case2}D2).
In Fig.~\ref{fig:case2}D3, he noticed that the borders around circles for ``\textit{OwnerAccount}" were blue~\bborder\, in calls of \textit{F0} but red~\rborder\, in calls of \textit{F1}, indicating that \textit{F0} increased it while \textit{F1} decreased it.
Furthermore, whenever \textit{F1} updated the ``\textit{OwnerAccount}" variable, it always reduced the value to zero, as evidenced by a red circle without an inner green circle~\redcircle\, showing the actual value (the green circle's radius was zero).
U1 found that the decrease caused by \textit{F1} (the width of the red border) following the same changing pattern as the value sent to the owner (the width of the blue curve for \textit{F1} in Fig.~\ref{fig:case2}D4), so he hypothesized that the ``\textit{OwnerAccount}" variable represented the value that the owner could withdraw via \textit{F1}.
Due to that this value only can be increased when the user invested by \textit{F0} and reduced to zero after the owner withdrew via \textit{F1}, U1 concluded that the owner could only withdraw a limited value of cryptocurrency from the contract.

After analyzing this smart contract with \techName{}, \mod{U1 deduced that 
this smart contract has a higher possibility of being a fraudulent smart contract, as evidenced by the negative balances of all users, the chain-like transactions, and the function allowing the owner to withdraw the investments. Such kind of patterns are commonly seen in fraudulent smart contracts~\cite{chen2021sadponzi}.
}
In fact, this contract is a typical Ponzi scheme called \textit{Suicide Watch}~\cite{suicidewatch}, and \techName{} reveals its risk intuitively without showing the complex source code.

\section{User Interview} \label{sec:interview}
We conducted semi-structured hour-long user interviews with 12 investors to evaluate the effectiveness and usability of \techName{}.

\subsection{Participants and Apparatus}
We recruited 12 participants (\textbf{U1-U12}) from universities and Web3 communities for our user interviews (5 females, 7 males, $age_{mean}= 27$, $age_{sd}= 3.67$). 
All participants have normal vision and are not color-blind.
They all have over one year of experience investing in smart contract applications.
Since our target users, investors, have different programming backgrounds in real-world scenarios, we evaluate the performance of \techName{} on participants both with and without the ability to read the smart contract source code.
Six participants (\textbf{U1-U6}) have the basic programming capability and are \textit{able} to read the smart contract source code (marked as \textit{\textbf{Investors-A}}), and the other six participants (\textbf{U7-U12}) do not have any technique background and are \textit{unable} to read the code (marked as \textit{\textbf{Investors-U}})).
Specifically, \textbf{U1-U4} are researchers in smart contract analysis and \mod{\textbf{U5, U6}} are smart contract programmers in a blockchain security company.
\textbf{U7-U12} are common investors participating in various Web3 projects.
Among them, \textbf{U7-U9} have blockchain-related work experiences, such as a product manager intern. 
Our interviews were online via Zoom. We launched \techName{} on the server and allowed participants to assess it via their
own laptops or desktops, sharing their screens with us.

\subsection{Procedure}
\begin{figure*}
    \centering
      \setlength{\abovecaptionskip}{0.2cm}
  \setlength{\belowcaptionskip}{-0.6cm}
    \includegraphics[width=0.7\linewidth]{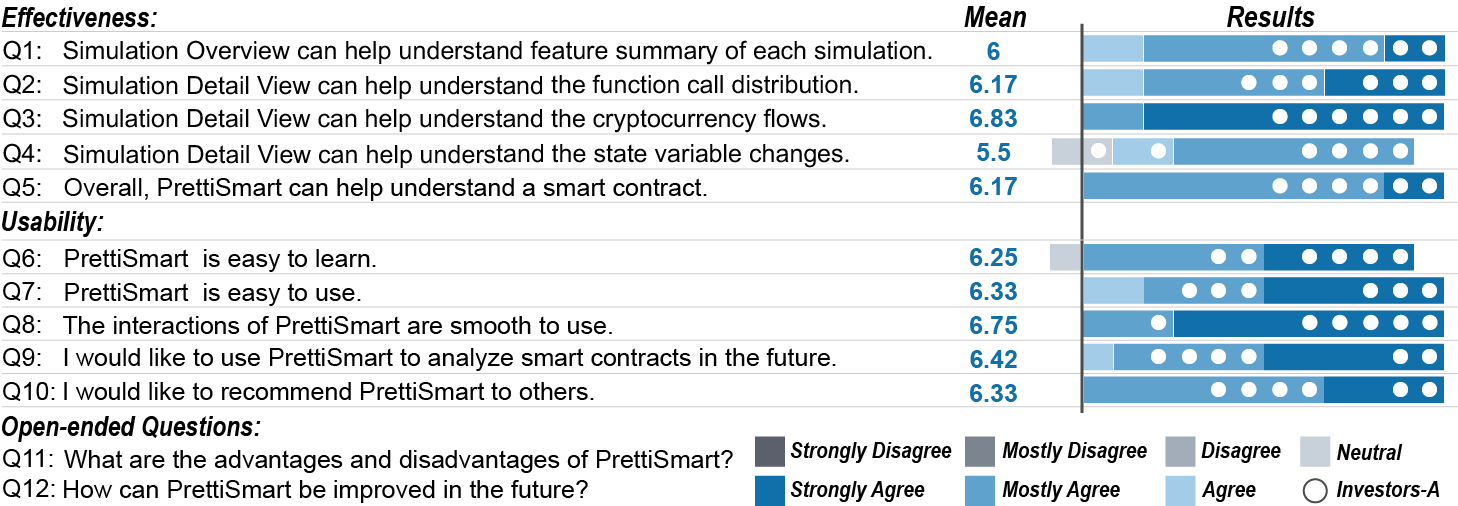}
    \caption{The user interview questionnaire results. Q1-Q11 are closed-ended rated on a 7-point Likert scale. \mod{Q11, Q12} are open-ended questions to collect participants' feedback. The detailed scores of Q1-Q10 are shown in a stacked bar chart, with \textit{\textbf{Investors-A}} marked by white circles.}
    \label{fig:result}
\end{figure*}

In the interview, we first asked participants to visit the online system of \techName{}, then introduced the background, visual design, interactions, and workflow of \techName{}.
Next, we demonstrated a usage scenario to guide participants on how to use \techName{} for analyzing a smart contract.
This tutorial phase lasted about 15 minutes.
In the following task phase, we first selected 12 smart contracts that can accept investments from previous smart contract analysis papers~\cite{durieux2020empirical,beillahi2020behavioral}, and then asked participants to use \techName{} to analyze one of these smart contracts.
Participants were asked to describe the behaviors of the smart contract in a think-aloud manner and provide the respective evidence they saw in \techName{}.
This task phase lasted until participants fully understood the smart contract, which usually lasted 30 minutes in our interviews.
Next, we invited participants to finish a post-study questionnaire with twelve questions (Q1-Q12), as shown in Fig.~\ref{fig:result}. Q1-Q10 are close-ended questions that should be answered on a 7-point Likert scale and are designed to measure \techName{}'s effectiveness and usability, following the question design of prior studies~\cite{li2021visual,wu2022defence} and \textit{PSSUQ} (Post-study System Usability Questionnaire)~\cite{lewis1992psychometric}.
\mod{Q11, Q12} are open-ended questions to collect participants' feedback on the shortcomings and improvement suggestions of \techName{}.
Overall, the user interview session for each participant took about 60 minutes.
We recorded the participants' data anonymously with their permission.

\subsection{Results}
Fig.~\ref{fig:result} summarizes the responses of participants to our post-study questionnaire and marks the responses from the six \textit{\textbf{Investors-A}}, capable of reading the source code, with white circles.
Overall, \techName{} received high ratings from the participants, with most scores for the closed-ended questions in the questionnaire being positive.
All the participants praised that compared with the source code, \techName{} provided a more intuitive and convincing interpretation of the functions and there were no existing tools designed for this.
However, there were three neutral scores for Q4 and Q6, which will be discussed later. 
The detailed feedback from participants can be summarized as follows:

\textbf{Effectiveness:}
According to scores of Q1-Q5, all participants recognized \techName{}'s visual designs and workflow.
For the \mO{} (Q1), participants agreed that using color to encode the balance changes helped them easily identify profit or loss patterns and understand the contract's purpose. For instance, most user balances displayed in red could indicate insufficient profitability or potential fraud.
Participants, especially \textit{\textbf{Investors-U}}, appreciated the design linking cryptocurrency flows (Q1) with function call distribution (Q2), as it intuitively revealed the outcomes of function calls without needing to review complex code.
\textit{\textbf{Investors-A}} U2 and U5 emphasized that this design demonstrated the economic model of the contract, while through the source code, they can only analyze the logic of individual functions, rather than the scenarios of multiple addresses' interactions.
For State Variable Changes (Q4), \textit{\textbf{Investors-A}} found it helpful in confirming the execution logic of smart contract functions. However, it was challenging for \textit{\textbf{Investors-U}} lacking programming experience.
U10 gave a neutral score, stating, ``I almost understand the smart contract's behaviors by analyzing the function call sequence and the cryptocurrency flows, but hardly find out how the state changes affect transactions". Despite this, \textit{\textbf{Investors-U}} U8 and U9 appreciated the effectiveness of State Variable Changes in verifying how the contract controls cryptocurrency flows, particularly when variable names were clear and when the cryptocurrency flows and state variables had similar changing trends. 
U3 gave another neutral score, mentioning that while the Variable Changes effectively showed the trends caused by functions, they lacked the source code details needed to fully understand the function execution logic. To address this, we plan to improve \techName{} by offering adjustable detail levels about state variables in future work.


\textbf{Usability:}
Overall, participants agreed that \techName{} was easy to learn and use, as shown in Fig.~\ref{fig:result} (\mod{Q6, Q7}).
U8 scored ``5" for Q6 and explained that it was easy for her to understand the visual encoding of investment-related concepts, such as cryptocurrency flows. 
However, she needed more time to learn the various types of state variables and the relations between their changes and the transaction patterns.
The scores for Q8 were all positive, indicating the current interactions were smooth for participants.
For Q10 and Q11, all the participants would like to use \techName{} and recommend it to other investors in the future.


\textbf{Limitations and improvements:}
\mod{
\techName{} is not without limitations.
First, scalability issues can arise.
For example, the number of simulations and function calls often increases with more simulated user and code complexity~\cite{nguyen2020sfuzz}.
This may result in scalability issues due to the limited space.
Following U4's suggestion, we added horizontal scroll bars in the simulation and simulation detail views to mitigate this issue.
In future work, we can develop an algorithm to automatically select highly representative simulations and function calls to reduce data scales.
Also, incorporating three layers in the \textit{function call details} may cause occlusions, such as the overlap between function calls and cryptocurrency flows. 
We can address this by adjusting opacity and allowing users to hide certain layers interactively.
Second, the names of functions and variables, defined in the source code, could affect users’ understanding of smart contracts.
Although \techName{} can help understand the underlying meaning of functions and variables by revealing variable changes and cryptocurrency flows after each function call, clear and meaningful names, like ``OwnerWithdraw", could help users quickly understand the smart contract's behavior. In future work, it is worth further exploration to automatically parse their semantic meaning from the code and rename them via large language models, further enhancing the usability of \techName{}. 
Additionally, due to the complexity and diversity of smart contracts, \techName{} can not guarantee the accuracy of user-drawn conclusions. For instance, a legitimate fundraising contract and a fraudulent one may exhibit similar behaviors, such as granting the owner control over user funds.
}
\section{Discussion}
In this section, we discuss the lessons learned during the development and evaluation of \techName{}, and its generalizability.

\textbf{Different perspectives from \textit{\textbf{Investors-A}} and \textit{\textbf{Investors-U}}.}
Despite having different backgrounds, both types of investors expressed strong appreciation for \techName{}. However, their purposes and ways of using \techName{} varied. 
Currently, \textit{\textbf{Investors-U}} often rely on auditing reports and hardly analyze contracts independently.
\techName{} addresses this gap by providing \textit{\textbf{Investors-U}} with the ability to intuitively understand the contract functionality, possible behaviors, and investment risk.
\textit{\textbf{Investors-U}} primarily relied on visual patterns about balance changes and cryptocurrency flows to understand the smart contract's behavior. 
Different from \textit{\textbf{Investors-U}}, \textit{\textbf{Investors-A}} may have the ability to understand each function from the source code, but they may struggle to foresee the contract's possible behaviors under various function calls from multiple users without actual transactions on the blockchain.
For \textit{\textbf{Investors-A}}, \techName{} serves not only as a tool for interpreting contracts but also for early verifying whether its usage aligns with expectations.
Specifically, they usually tended to understand the smart contracts by scrutinizing the changes made by each function.
%
%
%
%
%

\textbf{Towards better simulation for real-world scenarios.}
In this paper, we develop a simulator based on smart contract fuzzing, aiming to simulate the various function call patterns among multiple smart contract users to cover as many source code branches as possible.
While this simulation strategy ensures comprehensive coverage of function call scenarios, it may differ from the function call distribution in the real world, as indicated by feedback from our interviews.
For instance, even though owners of fraudulent smart contracts have the opportunity to invest in their own contracts like common investors, they are unlikely to do so and intentionally lose money.
Therefore, the simulator should generate the function call sequences that match the real-world behaviors as closely as possible.
Considering that smart contract users' behaviors are uncontrollable, we propose two possible directions to address this challenge: 
\textbf{(1)} learning user behavior from similar contracts with actual transactions, which is constrained by the availability of similar contracts~\cite{beillahi2020behavioral}.
\textbf{(2) }utilizing large language models (LLMs) to parse semantic meanings of functions and then modeling users as agents with different behavior patterns to simulate human behavior~\cite{park2023generative}.

\textbf{Generalizability.}
As the initial attempt to visually interpret smart contracts without showing source code, \techName{} focuses on the typical scenario of smart contract usage, i.e., a single smart contract transferring native tokens (e.g., the Ether).
With technology developing rapidly, smart contracts support transferring multiple tokens, such as ERC-20 tokens~\cite{victor2019measuring}, and a project can involve multiple smart contracts as well.
Since we regard the smart contract as a black box, the multiple contract scenarios have little impact on our visual designs but increase the complexity of the simulation.
For multiple tokens, the current cryptocurrency flow design can be used to show the total value of multiple tokens, and we have to make more efforts on differentiate tokens to generalize \techName{} to broaden scenarios.


\section{Conclusion}
We proposed a novel visualization approach, \techName{}, to provide cryptocurrency investors with an intuitive and reliable visual interpretation of smart contract via simulation.
Specifically, we first implemented a simulator to simulate the possible function call sequences in real-world scenarios.
Then, we proposed the visualization approach, \techName{}, to help investors understand the smart contract's behaviours after each function call.
We conducted two case studies and in-depth user interviews with 12 investors to evaluate the effectiveness and usability of \techName{}.
The results indicates that \techName{} can help investors intuitively understand smart contracts.

In future work, we plan to improve the simulator to better aligned with the real-world scenarios, and incorporate more interactions in \techName{} to satisfy investors with various backgrounds, such as manually modifying the function call sequences and specifying the risk preference of simulated users.
Also, since it is gradually more popular to use mobile devices for cryptocurrency investments~\cite{mirza2022mobile},
we plan to extend \techName{} to work on mobile devices.

\section*{Acknowledgments}
This project is supported by the Ministry of Education, Singapore, under its Academic Research Fund Tier 2 (Proposal ID: T2EP20222-0049).

\bibliographystyle{abbrv-doi-hyperref}

\bibliography{main}







\newpage
\onecolumn
\appendix
\section{Design Rationales and alternative designs}
In this section, we discuss the decision-making process behind our visual design and explore alternative design options, which are not described in the main paper due to the page limits.

For the Simulation Overview Module, we provided a barcode-based design that provides a visual summary for each simulation, addressing user needs for an overview of balance changes and unique functions involved. The barcode-based design, utilizing a gradient color scheme from red to blue to represent net balance after each function call, offers an intuitive, scalable solution with less visual clutter compared to line and area charts. 
Additionally, the circular markers for functions provide a cohesive overview of the Function Summary in the Simulation Detail Module, with unique function numbers and type-aligned colors consistent with the Function Summary.
For the used color for payable and non-payable functions,
we did consider applying two distinct color hues to represent non-payable and payable functions because using distinct colors for binary attributes is beneficial.
However, this would introduce many different color hues into one single view (simulation detail view) as well as the whole system, potentially causing confusion and making it hard to differentiate between encodings for different information types, such as balances and state variables.
After considering all these factors and the priority of different data to be visualized, we decided to use the current color scheme to indicate payable and non-payable, and we have increased the contrast between the light and dark colors.

The Function Summary is designed to provide an overview of each function's behaviors, with a focus on the cryptocurrency flows triggered by function calls. 
It considered three flow directions (i.e., to the contract, to the caller, and to others) that could be caused by function calls.
Upward triangles indicate ``\textit{to the contract}", while downward triangles inside and outside function rectangles indicate ``\textit{to the caller}" and ``\textit{to others}", respectively.
The use of triangles, informed by feedback from the preliminary study, reflects experts' suggestions to highlight the crypto flow directions with intuitive arrow-like symbols pointing to users or the contract. 
Our design maintains the design intuition and further differentiates between the flows from other users and the flows from the caller itself by the positions of the downward triangles.
Additionally, the three bars on the right display the proportion of cryptocurrency flow to the contract and users, offering statistical context for different flow types.

Function Call Details focuses on integrating multiple pieces of information to help users analyze function call sequences and their associated behaviors, including balance changes, cryptocurrency flows, and state variable changes. The Massive Sequence Views (MSV)[1] have been proven effective for visualizing temporal (net balance changes) and structural patterns (cryptocurrency flows) of dynamic networks, and have been used for visualizing cryptocurrency flows [2].
Its timeline-based design also makes it intuitive to incorporate the event sequential data like function calls.
Therefore, we use MSV as the foundation layout for the Function Call Details and further refine the visual design to meet other user requirements.
Specifically, we replace the vertical lines in MSV with two types of curves in distinct shapes and colors to enhance users' intuition about flow directions.
The function call sequences are superimposed to facilitate the analysis of relationships between cryptocurrency flows and function call behaviors.
The area charts are embedded above and below the bar area of each address's function call sequence to highlight net balance changes over time and avoid visual clutter.
Given the uncertain number of state variables, incorporating variable changes into the above design is challenging. To address this, we present detailed state variable changes below the aforementioned design, aligned with the same timeline. Based on feedback from the preliminary study, we use two shapes to encode numerical data and addresses, as well as highlight changes and periods of stability.

~\\
\noindent[1] S. v. d. Elzen, D. Holten, J. Blaas and J. J. van Wijk, "Dynamic Network Visualization withExtended Massive Sequence Views," in IEEE Transactions on Visualization and Computer Graphics, vol. 20, no. 8, pp. 1087-1099, 2013.

\noindent[2] X. Wen, Y. Wang, X. Yue, F. Zhu, and M. Zhu. Nftdisk: Visual detection
of wash trading in nft markets. In Proceedings of the 2023 CHI Conference
on Human Factors in Computing Systems, pp. 1–15, 2023. 3 
\end{document}